\documentclass{llncs}
\usepackage{times}
\usepackage{abbrev}
\usepackage{trackchanges}

\usepackage{listings}

\usepackage[utf8]{inputenc} 
\usepackage[T1]{fontenc}

\usepackage{amsmath}
\usepackage{amsfonts}
\usepackage{amssymb}
\usepackage{a4,a4wide}
\author{Alexandre Miguel Pinto \and Lu\'is Moniz Pereira \\
		\{amp$|$lmp\}@di.fct.unl.pt}
\title{Each \nlp\ has a \twov\\ \MHs}
\institute{Centro de Intelig\^encia Artificial (CENTRIA), Departamento de Inform\'atica\\
		Faculdade de Ci\^encias e Tecnologia, Universidade Nova de Lisboa\\
		2829-516 Caparica, Portugal}

\begin{document}
	\maketitle
	\begin{abstract}
		In this paper we explore a unifying approach --- that of hypotheses assumption --- as a means to provide a semantics for all \NLPs\ 
		(NLPs), the \MH\ (MH) semantics
		\footnote{	This paper is a very condensed summary of some of the main contributions of the PhD Thesis \cite{PHDAMP} of the first author, supported by FCT-MCTES grant SFRH / BD / 28761 / 2006, and supervised by the second author.}.
		This semantics takes a positive hypotheses assumption approach as a means to guarantee the desirable properties of \m\ existence,
		relevance and cumulativity, and of generalizing the \SMss\ in the process.
		To do so we first introduce the fundamental semantic concept of minimality of assumed positive hypotheses, define the MH semantics, and
		analyze the semantics' properties and applicability. 
		Indeed, abductive Logic Programming can be conceptually captured by a strategy centered on the assumption of abducibles (or hypotheses).
		Likewise, the Argumentation perspective of \LPs\ (e.g. \cite{dung95acceptability}) also lends itself to an arguments (or hypotheses)
		assumption approach.
		Previous works on Abduction (e.g. \cite{DBLP:journals/logcom/KakasKT92}) have depicted the atoms of default
		negated literals in NLPs as abducibles, i.e., assumable hypotheses.
		We take a complementary and more general view than these works to NLP semantics by employing positive hypotheses instead.		
		
	\end{abstract}
	
	\section{Background}
		\LPs\ have long been used in \KRR.
		\mydef{\NLP}{nlp}{
			By an alphabet $\mathcal{A}$ of a language $\mathcal{L}$ we mean (finite or countably infinite) disjoint sets of constants, predicate 
			symbols, and function symbols, with at least one constant.
			In addition, any alphabet is assumed to contain a countably infinite set of distinguished variable symbols.
			A term over $\mathcal{A}$ is defined recursively as either a variable, a constant or an expression of the form $f(t_1, \ldots, t_n)$ where 
			$f$ is a function symbol of $\mathcal{A}$, $n$ its arity, and the $t_i$ are terms.
			An atom over $\mathcal{A}$ is an expression of the form $P(t_1, \ldots, t_n)$ where $P$ is a predicate symbol of $\mathcal{A}$, and
			the $t_i$ are terms.
			A literal is either an atom $A$ or its default negation $\dnot A$.
			We dub default literals (or default negated literals --- DNLs, for short) those of the form $\dnot A$.
			A term (resp. atom, literal) is said ground if it does not contain variables.
			The set of all ground terms (resp. atoms) of $\mathcal{A}$ is called the Herbrand universe (resp. base) of $\mathcal{A}$.
			For short we use $\mathcal{H}$ to denote the Herbrand base of $\mathcal{A}$.				
			A \NLP\ (NLP) is a possibly infinite set of rules (with no infinite descending chains of syntactical dependency) of the form
			$$ H \<  B_1, \ldots, B_n, \dnot C_1, \ldots, \dnot C_m, \textrm{ (with }m,n\geq 0\textrm{ and finite)} $$
			where $H$, the $B_i$ and the $C_j$ are atoms, and each rule stands for all its ground instances.
			In conformity with the standard convention, we write rules of the form $H \leftarrow$ also simply as $H$ (known as ``facts'').
			An NLP $P$ is called definite if none of its rules contain default literals.
			$H$ is the head of the rule $r$, denoted by $head(r)$, and $body(r)$ denotes the set
			$\{B_1, \ldots, B_n, \dnot C_1, \ldots, \dnot C_m\}$ of all the literals in the body of $r$.
		}
		When doing problem modelling with \lps, rules of the form
		$$ \bot\<  B_1, \ldots, B_n, \dnot C_1, \ldots, \dnot C_m, \textrm{ (with }m,n\geq 0\textrm{ and finite)} $$
		with a non-empty body are known as a type of \ICs\ (ICs), specifically \emph{denials}, and they are normally used to prune out
		unwanted candidate solutions.
			We abuse the `$\dnot$' default negation notation applying it to non-empty sets of literals too: we write $\dnot S$ to denote 
		$\{\dnot s:s\in S\}$, and confound $\dnot\dnot a\equiv a$.
		When $S$ is an arbitrary, non-empty set of literals $S=\{B_1, \ldots, B_n, \dnot C_1, \ldots, \dnot C_m\}$ we use 
		\begin{itemize}
			\item $S^{+}$ denotes the set $\{B_1, \ldots, B_n\}$ of positive literals in $S$
			\item $S^{-}$ denotes the set $\{\dnot C_1, \ldots, \dnot C_m\}$ of negative literals in $S$
			\item $|S|=S^{+}\cup(\dnot S^{-})$ denotes the set $\{B_1, \ldots, B_n, C_1, \ldots, C_m\}$ of atoms of $S$
		\end{itemize}
		As expected, we say a set of literals $S$ is consistent iff $S^{+}\cap|S^{-}|=\emptyset$.
		We also write $heads(P)$ to denote the set of heads of non-IC rules of a (possibly constrained) program $P$, i.e., 
		$heads(P)=\{head(r):r\in P\}\setminus\{\bot\}$, and $facts(P)$ to denote the set of facts of $P$ --- 
		$facts(P)=\{head(r):r\in P\wedge body(r)=\emptyset\}$.
		\mydef{Part of body of a rule not in loop}{overlineBody}{
			Let $P$ be an NLP and $r$ a rule of $P$.
			We write $\overline{body(r)}$ to denote the subset of $body(r)$ whose atoms do not depend on $r$.
			Formally, $\overline{body(r)}$ is the largest set of literals such that
			$$\overline{body(r)}\subseteq body(r) \wedge 
				\forall_{a\in |\overline{body(r)}|}\nexists_{r_a\in P}(head(r_a)=a\wedge r_a\twoheadleftarrow r)$$
			where $r_a\twoheadleftarrow r$ means rule $r_a$ depends on rule $r$, i.e., either $head(r)\in|body(r_a)|$ or
			there is some other rule $r'\in P$ such that $r_a\twoheadleftarrow r'$ and $head(r)\in|body(r')|$.
		}

		\mydef{\LSed\ and Classically supported interpretations}{lsupport}{
			We say an interpretation $I$ of an NLP $P$ is layer (classically) supported iff every atom $a$ of $I$ is layer (classically) supported in $I$.
			$a$ is layer (classically) supported in $I$ iff there is some rule $r$ in $P$ with $head(r)=a$ such that $I\models\overline{body(r)}$ ($I\models body(r)$).
			Likewise, we say the rule $r$ is layer (classically) supported in $I$ iff $I\models\overline{body(r)}$ ($I\models body(r)$).
		}
		
		Literals in $\overline{body(r)}$ are, by definition, not in loop with $r$.
		The notion of \ls\ requires that all such literals be \true\ under $I$ in order for $head(r)$ to be \lsed\ in $I$.
		Hence, if $\overline{body(r)}$ is empty, $head(r)$ is \emph{ipso facto} \lsed.
		\myproposition{Classical Support implies Layered Support}{cs=>ls}{
			Given a NLP $P$, an interpretation $I$, and an atom $a$ such that $a\in I$, if $a$ is classically supported in $I$ then $a$ is also \lsed\
			in $I$.
		}{
			Knowing that, by definition, $\overline{body(r)}\subseteq body(r)$ for every rule $r$, it follows trivially that $a$ is \lsed\ in $I$ if $a$ is
			classically supported in $I$.
		}		
	\section{Motivation}
		``Why the need for another 2-valued semantics for NLPs since we already have the \SMs\ one?''
		The question has its merit since the \SMs\ (SMs) semantics \cite{SM-GL} is exactly what is necessary for so many problem solving issues, but the answer to it is best 
		understood when we ask it the other way around:
		``Is there any situation where the SMs semantics does not provide all the intended \ms?'' and
		``Is there any \twov\ generalization of SMs that keeps the intended \ms\ it does provide, adds the missing intended ones, and also enjoys
		the useful properties of guarantee of \m\ existence, relevance, and cumulativity?''
	
			\myex{A Joint Vacation Problem --- Merging \LPs}{vacation}{
			Three friends are planning a joint vacation.
			First friend says ``If we don't go to the mountains, then we should go to the beach''.
			The second friend says ``If we don't go to travelling, then we should go to the mountains''.
			The third friend says ``If we don't go to the beach, then we should go travelling''.
			We code this information as the following NLP:
			\myprog{
				beach & \< & \dnot mountain\\
				mountain & \< & \dnot travel\\
				travel & \< & \dnot beach
			}
			Each of these individual consistent rules come from a different friend.
			According to the SMs, each friend had a ``solution'' (a SM) for his own rule, but when we put the three rules together, because they
			form an \olon\ (OLON), the resulting merged \lp\ has no SM.
			If we assume $beach$ is \true\ then we cannot conclude $travel$ and therefore we conclude $mountain$ is also \true\ ---
			this gives rise to the $\{beach,mountain,\dnot travel\}$ joint and multi-place vacation solution.
			The other (symmetric) two are $\{mountain,\dnot beach, travel\}$ and $\{travel,\dnot mountain,\\ beach\}$.
			This example too shows the importance of having a \twovs\ guaranteeing \m\ existence, in this case for the sake of arbitrary merging of
			\lps\ (and for the sake of existence of a joint vacation for these three friends).
		}
		\paragraph{{\bf Increased Declarativity.}}\label{subsec:increasedDeclarativity}
			An IC is a rule whose head is $\bot$, and although such syntactical definition of IC is generally accepted as standard, the
			SM semantics can employ \olons, such as the $a\<\dnot a,X$ to act as ICs, thereby mixing and unnecessarily confounding two distinct
			\KR\ issues: the one of IC use, and the one of assigning semantics to loops.
			For the sake of declarativity, rules with $\bot$ head should be the only way to write ICs in a LP: no rule, 
			or combination of rules, with head different from $\bot$ should possibly act as IC(s) under any given semantics.
			\amp{
			It is commonly argued that answer sets (or \sms) of a program correspond to the solutions of the corresponding problem, so no answer
			set means no solution.
			We argue against this position: ``normal'' logic rules (i.e., non-ICs) should be used to shape the candidate-solution space, whereas
			ICs, and ICs alone, should be allowed to play the role of cutting down the undesired candidate-solutions.
			In this regard, an IC-free NLP should always have a \m; if some problem modelled as an NLP with ICs has no solution (i.e., no \m) that
			should be due only to the ICs, not to the ``normal'' rules.
			}
		\paragraph{{\bf Argumentation}}\label{subsec:modelingArgumentation}
			From an argumentation perspective, the author of \cite{dung95acceptability}, states:
			\begin{quotation}
				{\em ``Stable extensions do not capture the intuitive semantics of every meaningful argumentation system.''}
			\end{quotation}
			where the ``stable extensions'' have a 1-to-1 correspondence to the SMs (\cite{dung95acceptability}), and also
			\begin{quotation}
				{\em ``Let $P$ be a knowledge base represented either as a logic program, or as a nonmonotonic theory or as an argumentation 
						framework. 
						Then there is not necessarily a ``bug'' in $P$ if $P$ has no stable semantics.
						
						This theorem defeats an often held opinion in the logic programming and nonmonotonic reasoning community that if a \lp\ 
						or a nonmonotonic theory has no stable semantics then there is something ``wrong'' in it.''}
			\end{quotation}
			Thus, a criterion different from the \emph{stability} one must be used in order to effectively \m\ every argumentation framework adequately.
		\paragraph{{\bf Arbitrary Updates and/or Merges}}\label{subsec:allowingArbitraryUpdatesMerges}
			One of the main goals behind the conception of non-monotonic logics was the ability to deal with the changing, evolving, updating of 
			knowledge.
			There are scenarios where it is possible and useful to combine several Knowledge Bases (possibly from different authors or sources) into a 
			single one, 
			and/or to update a given KB with new knowledge. 
			Assuming the KBs are coded as IC-free NLPs, as well as the updates, the resulting KB is also an IC-free NLP.
			In such a case, the resulting (merged and/or updated) KB should always have a semantics.
			This should be true particularly in the case of NLPs where no negations are allowed in the heads of rules.
			In this case no contradictions can arise because there are no conflicting rule heads.
			The lack of such guarantee when the underlying semantics used is the \SMs, for example, compromises the possibility of arbitrarily
			updating and/or merging KBs (coded as IC-free NLPs).
			In the case of self-updating programs, the desirable ``liveness'' property is put into question, even without outside intervention.
	
			These motivational issues raise the questions
			``Which should be the \twov\ \ms\ of an NLP when it has no SMs?'',
			``How do these relate to SMs?'',
			``Is there a uniform approach to characterize both such \ms\ and the SMs?'', and 
			``Is there any \twov\ generalization of the SMs that encompasses the intuitive semantics of \emph{every} 
			\lp?''.
			Answering such questions is a paramount motivation and thrust in this paper.
		\subsection{Intuitively Desired Semantics}\label{subsec:intuitivelydesired}
			It is commonly accepted that the non-stratification of the default $\dnot$ is the fundamental ingredient which allows for the possibility of 
			existence of several \ms\ for a program. 
			The non-stratified DNLs (i.e., in a loop) of a program can thus be seen as non-deterministically assumable choices.
			The rules in the program, as well as the particular semantics we wish to assign them, is what constrains which sets of those choices
			we take as acceptable.
			Programs with OLONs (ex. \ref{ex:vacation}) are said to be
			``contradictory'' by the SMs community because the latter takes a negative hypotheses assumption approach,
			consistently maximizing them,
			i.e., DNLs are seen as assumable/abducible hypotheses.
			In ex.\ref{ex:vacation} though, assuming whichever maximal negative hypotheses leads to a positive contradictory
			conclusion via the rules.
			On the other hand, if we take a consistent minimal \emph{positive} hypotheses assumption (where the assumed
			hypotheses are the \emph{atoms} of the DNLs), then it is impossible to achieve a
			contradiction since no negative conclusions can be drawn from NLP rules.
			Minimizing positive assumptions implies the maximizing of negative ones but gaining an extra degree of freedom.

		\subsection{Desirable Formal Properties}\label{subsec:desirableproperties}
			Only ICs (rules with $\bot$ head) should ``endanger'' \m\ existence in a \lp.
			Therefore, a semantics for NLPs with no ICs should guarantee \m\ existence (which, e.g., does not occur with SMs).
			Relevance is also a useful property since it allows the development of top-down query-driven proof-procedures that allow for the
			sound and complete search for answers to a user's query.
			This is useful in the sense that in order to find an answer to a query only the relevant part of the program must be considered, whereas
			with a non-relevant semantics 
			the whole program must be considered, with corresponding performance disadvantage compared
			to a relevant semantics.
			
			\mydef{Relevant part of $P$ for atom $a$}{Rel_P(a)}{
				The relevant part of NLP $P$ for atom $a$ is\\
				$Rel_P(a)=\{r_a\in P:head(r_a)=a\}\cup\{r\in P:\exists_{r_a\in P\wedge head(r_a)=a}r_a\twoheadleftarrow r\}$
			}
			
			\mydef{Relevance (adapted from \cite{dix93aclassificationtheoryii})}{relevance}{
				A semantics $Sem$ for \lps\ is said Relevant iff for every program $P$
				$$\forall_{a\in\mathcal{H}_{P}}
					(\forall_{M\in Models_{Sem}(P)}a \in M) \Leftrightarrow 
					(\forall_{M_{a}\in Models_{Sem}(Rel_{P}(a))}a \in M_{a})$$
			}			
			Moreover, cumulativity also plays a role in performance enhancement in the sense that only a semantics enjoying this property can take 
			advantage of storing intermediate lemmas to speed up future computations.
			\mydef{Cumulativity (adapted from \cite{Dix95aclassification})}{cumulativity}{
				Let $P$ be an NLP, and $a,b$ two atoms of $\mathcal{H}_{P}$.
				A semantics $Sem$ is Cumulative iff the semantics of $P$ remains unchanged when any atom \true\ in the semantics is 
				added to $P$ as a fact:
				$$ \forall_{a,b\in\mathcal{H}_{P}}\big((\forall_{M\in Models_{Sem}(P)} a\in M)\Rightarrow$$
				$$(\forall_{M\in Models_{Sem}(P)} b\in M\Leftrightarrow 
					\forall_{M_{a}\in Models_{Sem}(P\cup\{a\})} b\in M_{a})\big)$$
			}
			Finally, each individual SM of a program, by being minimal and classically supported, should be accepted as a \m\ according to
			every \twovs, and hence every \twovs\ should be a \m\ conservative extension of \SMs.
	\section{Syntactic Transformations}
		It is commonly accepted that definite LPs (i.e., without default negation) have only one \twov\ \m\ --- its \emph{least \m} which coincides
		with the \WFM\ (WFM \cite{WFS}).
		This is also the case for locally stratified LPs.
		In such cases we can use a syntactic transformation on a program to obtain that \m.
		In \cite{DBLP:journals/tplp/BrassDFZ01} the author defined the program Remainder (denoted by $\widehat{P}$) for calculating the
		WFM, which coincides with the unique perfect \m\ for locally stratified LPs.
		The Remainder can thus be seen as a generalization for NLPs of the $lfp(T)$, the latter obtainable only from the subclass of definite LPs.
		We recap here the definitions necessary for the Remainder because we will use it in the definition of our \MHs.
		The intuitive gist of MH semantics (formally defined in section \ref{sec:MHs}) is as follows: an interpretation $M_H$ is a MH \m\ of program $P$ iff 
		there is some minimal set of hypotheses $H$ such that the truth-values of all atoms of $P$ become determined assuming the atoms in $H$
		as \true.
		We resort to the program Remainder as a deterministic (and efficient, i.e., computable in polynomial time) means to find out if
		the truth-values of all literals became determined or not --- we will see below how the Remainder can be used to find this out.
		
		\subsection{Program Remainder}
			For self-containment, we include here the definitions of \cite{DBLP:journals/tplp/BrassDFZ01} upon which the Remainder 
			relies, and adapt them where convenient to better match the syntactic conventions used throughout this paper.
			\mydef{Program transformation (def. 4.2 of \cite{DBLP:journals/tplp/BrassDFZ01})}{progTransf}{
				A \emph{program transformation} is a relation $\mapsto$ between ground \lps.
				A semantics $S$ allows a transformation $\mapsto$ iff $Models_{S}(P_{1})=Models_{S}(P_{2})$ for all $P_{1}$ and $P_{2}$ with 
				$P_{1}\mapsto P_{2}$.
				We write $\mapsto^{*}$ to denote the fixed point of the $\mapsto$ operation, i.e., 
				$ P\mapsto^{*}P'$ where $\nexists_{P''\neq  P'}P'\mapsto P''$.
				It follows that $P\mapsto^{*}P'\Rightarrow P'\mapsto P'$.
			}
			\mydef{Positive reduction (def. 4.6 of \cite{DBLP:journals/tplp/BrassDFZ01})}{posReduct}{
				Let $P_{1}$ and $P_{2}$ be ground programs.
				Program $P_{2}$ results from $P_{1}$ by \emph{positive reduction} $(P_{1}\mapsto_{P} P_{2})$ iff there is a rule $r\in P_{1}$ and a 
				negative literal $\dnot b\in body(r)$ such that $b\notin heads(P_{1})$, i.e., there is no rule for $b$ in $P_{1}$, and 
				$P_{2}=(P_{1}\setminus\{r\})\cup\{head(r)\<(body(r)\setminus\{\dnot b\})\}$.
			}			
			\mydef{Negative reduction (def. 4.7 of \cite{DBLP:journals/tplp/BrassDFZ01})}{negReduct}{
				Let $P_{1}$ and $P_{2}$ be ground programs.
				Program $P_{2}$ results from $P_{1}$ by \emph{negative reduction} $(P_{1}\mapsto_{N} P_{2})$ iff there is a rule $r\in P_{1}$ and a 
				negative literal $\dnot b\in body(r)$ such that $b\in facts(P_{1})$, i.e., $b$ appears as a fact in $P_{1}$, and 
				$P_{2}=P_{1}\setminus\{r\}$.
			}				
			Negative reduction is consistent with classical support, but not with the layered one.
			Therefore, we introduce now a layered version of the negative reduction operation.
			\mydef{Layered negative reduction}{layeredNegReduct}{
				Let $P_{1}$ and $P_{2}$ be ground programs.
				Program $P_{2}$ results from $P_{1}$ by \emph{layered negative reduction} $(P_{1}\mapsto_{LN} P_{2})$ iff there is a rule 
				$r\in P_{1}$ and a negative literal $\dnot b\in\overline{body(r)}$ such that $b\in facts(P_{1})$, i.e., $b$ appears as a fact in $P_{1}$, 
				and $P_{2}=P_{1}\setminus\{r\}$.
			}	
			The \SCCs\ (SCCs) of rules of a program can be calculated in polynomial time \cite{citeulike:2204443}.
			Once the SCCs of rules have been identified, the $\overline{body(r)}$ subset of $body(r)$, for each rule $r$, is
			identifiable in linear time --- one needs to check just once for each literal in $body(r)$ if it is also in $\overline{body(r)}$.
			Therefore, these polynomial time complexity operations are all the added complexity Layered negative reduction adds over regular
			Negative reduction.
			\mydef{Success (def. 5.2 of \cite{DBLP:journals/tplp/BrassDFZ01})}{success}{
				Let $P_{1}$ and $P_{2}$ be ground programs.
				Program $P_{2}$ results from $P_{1}$ by \emph{success} $(P_{1}\mapsto_{S} P_{2})$ iff there are a rule $r\in P_{1}$ and a fact
				$b\in facts(P_{1})$ such that $b\in body(r)$, and $P_{2}=(P_{1}\setminus\{r\})\cup\{head(r)\<(body(r)\setminus\{b\})\}$.
			}
			\mydef{Failure (def. 5.3 of \cite{DBLP:journals/tplp/BrassDFZ01})}{failure}{
				Let $P_{1}$ and $P_{2}$ be ground programs.
				Program $P_{2}$ results from $P_{1}$ by \emph{failure} $(P_{1}\mapsto_{F} P_{2})$ iff there are a rule $r\in P_{1}$ and a positive 
				literal $b\in body(r)$ such that $b\notin heads(P_{1})$, i.e., there are no rules for $b$ in $P_{1}$, and $P_{2}=P_{1}\setminus\{r\}$.
			}
			\mydef{Loop detection (def. 5.10 of \cite{DBLP:journals/tplp/BrassDFZ01})}{loopDetection}{
				Let $P_{1}$ and $P_{2}$ be ground programs.
				Program $P_{2}$ results from $P_{1}$ by \emph{loop detection} $(P_{1}\mapsto_{L} P_{2})$ iff there is a set $\mathcal{A}$ of ground 
				atoms such that
				\begin{enumerate}
					\item for each rule $r\in P_{1}$, if $head(r)\in\mathcal{A}$, then $body(r)\cap\mathcal{A}\neq\emptyset$,
					\item $P_{2}:=\{r\in P_{1}|body(r)\cap\mathcal{A}=\emptyset\}$,
					\item $P_{1}\neq P_{2}$.
				\end{enumerate}		
			}
			We are not entering here into the details of the \emph{loop detection} step, but just taking note that
			1) such a set $\mathcal{A}$ corresponds to an unfounded set (cf. \cite{WFS}); 
			2) loop detection is computationally equivalent to finding the SCCs \cite{citeulike:2204443}, and is known to be of polynomial time
			complexity; and
			3) the atoms in the unfounded set $\mathcal{A}$ have all their corresponding rules involved in SCCs where all heads of rules in loop
			appear positive in the bodies of the rules in loop.
			\mydef{Reduction (def. 5.15 of \cite{DBLP:journals/tplp/BrassDFZ01})}{reduction}{\\
				Let $\mapsto_{X}$ denote the rewriting system:\hspace{5mm}
				$\mapsto_{X}:=\mapsto_{P}\cup\mapsto_{N}\cup\mapsto_{S}\cup\mapsto_{F}\cup\mapsto_{L}$.
			}
			\mydef{Layered reduction}{layeredReduction}{\\
				Let $\mapsto_{LX}$ denote the rewriting system:
				$\mapsto_{LX}:=\mapsto_{P}\cup\mapsto_{LN}\cup\mapsto_{S}\cup\mapsto_{F}\cup\mapsto_{L}$.
			}				
			\mydef{Remainder (def. 5.17 of \cite{DBLP:journals/tplp/BrassDFZ01})}{remainder}{
				Let $P$ be a program.
				Let $\widehat{P}$ satisfy\\
					$ground(P)\mapsto_{X}^{*}\widehat{P}$.
				Then $\widehat{P}$ is called the \emph{remainder} of $P$, and is guaranteed to exist and to be unique to $P$.
				Moreover, the calculus of $\mapsto_{X}^{*}$ is known to be of polynomial time complexity
				\cite{DBLP:journals/tplp/BrassDFZ01}.
				When convenient, we write $Rem(P)$ instead of $\widehat{P}$.
			}
			An important result from \cite{DBLP:journals/tplp/BrassDFZ01} is that the WFM of $P$ is such that
			$WFM^+(P)=facts(\widehat{P})$, $WFM^{+u}=heads(\widehat{P})$, and $WFM^-(P)=\mathcal{H}_P\setminus WFM^{+u}(P)$, 
			where 
			$WFM^+(P)$ denotes the set of atoms of $P$ \true\ in the WFM,
			$WFM^{+u}(P)$ denotes the set of atoms of $P$ \true\ or \undef\ in the WFM, and
			$WFM^-(P)$ denotes the set of atoms of $P$ \false\ in the WFM.
			\mydef{Layered Remainder}{layeredRemainder}{
				Let $P$ be a program.
				Let the program $\mathring{P}$ satisfy
					$ground(P)\mapsto_{LX}^{*}\mathring{P}$.
				Then $\mathring{P}$ is called a \emph{layered remainder} of $P$.
				Since $\mathring{P}$ is equivalent to $\widehat{P}$, apart from the difference between $\mapsto_{LN}$ and $\mapsto_{N}$,
				it is trivial that $\mathring{P}$ is also guaranteed to exist and to be unique for $P$.
				Moreover, the calculus of $\mapsto_{LX}^{*}$ is likewise of polynomial time complexity because $\mapsto_{LN}$ is also of
				polynomial time complexity.
			}
			The remainder's rewrite rules are provably confluent, ie. independent of application order.
			The layered remainder's rules differ only in the negative reduction rule and the confluence proof of the former is readily adapted to the latter.
			\myex{$\mathring{P}$ versus $\widehat{P}$}{LRem(P)vsRem(P)}{
				Recall the program from example \ref{ex:vacation} but now with an additional fourth stubborn friend who insists on going to the 
				beach no matter what.
				$P=$
				\myprog{
					beach & \< & \dnot mountain\\
					mountain & \< & \dnot travel\\
					travel & \< & \dnot beach\\
					beach &&
				}
				We can clearly see that the single fact rule does not depend on any other, and that the remaining three rules forming the loop all
				depend on each other and on the fact rule $beach$.
				$\widehat{P}$ is the fixed point of $\mapsto_{X}$, i.e., the fixed point of 
				$\mapsto_{P}\cup\mapsto_{N}\cup\mapsto_{S}\cup\mapsto_{F}\cup\mapsto_{L}$.
				Since $beach$ is a fact, the $\mapsto_{N}$ 
				transformation deletes the
				$travel\<\dnot beach$ rule;
				i.e., $P\mapsto_{N}P'$ is such that
				
				$P'=\{beach\<\dnot mountain$\ \ \ \ \ \ \ $mountain \<\dnot travel$\ \ \ \ \ \ \ $beach\<\}$
				
				Now in $P'$ there are no rules for $travel$ and hence we can apply the $\mapsto_{P}$ 
				transformation which deletes the $\dnot travel$ from the body of $mountain$'s rule; i.e, $P'\mapsto_{P}P''$
				where 
				$P''=\{beach\<\dnot mountain$\ \ \ \ \ \ \ $mountain\<$\ \ \ \ \ \ \ $beach\<\}$

				Finally, in $P''$ $mountain$ is a fact and hence we can again apply the $\mapsto_{N}$ obtaining $P''\mapsto_{P}P'''$ where 
				$P'''=\{mountain\<$\ \ \ $beach\<\}$
				upon which no more transformations can be applied, so $\widehat{P}=P'''$.
				Instead, $\mathring{P}=P$ is the fixed point of $\mapsto_{LX}$, i.e., the fixed point of 
				$\mapsto_{P}\cup\mapsto_{LN}\cup\mapsto_{S}\cup\mapsto_{F}\cup\mapsto_{L}$.
			}
	\section{\MHS}\label{sec:MHs}
		\subsection{Choosing Hypotheses}
			The abductive perspective of \cite{DBLP:journals/logcom/KakasKT92} depicts the atoms of DNLs as abducibles,
			i.e., assumable hypotheses.	
			Atoms of DNLs can be considered as abducibles, i.e., assumable hypotheses, but not all of them.
			When we have a locally stratified program we cannot really say there is any degree of freedom in
			assuming 	truth values for the atoms of the program's DNLs.
			So, we realize that only the atoms of DNLs involved in SCCs\footnote{\SCCs, as in Examples
			\ref{ex:vacation} and \ref{ex:LRem(P)vsRem(P)}}			
			are eligible to be considered further assumable
			hypotheses.
	
			Both the SMs and the approach of \cite{DBLP:journals/logcom/KakasKT92}, when taking the abductive perspective, adopt negative
			hypotheses only.
			This approach works fine for some instances of non-\wf\ negation such as loops (in particular, for \elons\ like this one), but not for
			\olons\ like, e.g. $a\<\dnot a$:
			assuming $\dnot a$ would lead to the conclusion that $a$ is \true\ which contradicts the initial assumption.
			To overcome this problem, we generalized the hypotheses assumption perspective to allow the adoption, not only of negative hypotheses,
			but also of positive ones.
			Having taken this generalization step we realized that positive hypotheses assumption alone is sufficient to address all situations, i.e.,
			there is no need for both positive and negative hypotheses assumption.
			Indeed, because we minimize the positive hypotheses we are with one stroke maximizing the negative ones, which has been the traditional
			way of dealing with the CWA, and also with \sms\ because the latter's requirement of classical support minimizes \ms.
			
			In example \ref{ex:vacation} we saw three solutions, each assuming as \true\ one of the DNLs in the loop.
			Adding a fourth stubborn friend insisting on going to the beach, as in example \ref{ex:LRem(P)vsRem(P)}, should still permit the two
			solutions $\{beach, mountain, \dnot travel\}$ and $\{travel,\dnot mountain, beach\}$.
			The only way to permit both these solutions is by resorting to the Layered Remainder, and not to the Remainder, as a means to identify
			the set of assumable hypotheses.
			
			Thus, all the literals of $P$ that are not determined \false\ in
			$\mathring{P}$
			are candidates for the role of hypotheses we may consider to assume as \true.
			Merging this perspective with the abductive perspective of \cite{DBLP:journals/logcom/KakasKT92} (where the DNLs are the abducibles) we 
			come to the following definition of the Hypotheses set of a program.
			
	%
			\mydef{Hypotheses set of a program}{hyps}{
				Let $P$ be an NLP.
				We write $Hyps(P)$ to denote the set of assumable hypotheses of $P$: the atoms that appear as DNLs in the bodies of 
				rules of $\mathring{P}$.
				Formally, $Hyps(P)=\{a:\exists_{r\in \mathring{P}}\dnot a\in body(r)\}$.
			}
			One can define a classical support compatible version of the Hypotheses set of a program, only using to that effect the Remainder 
			instead of the Layered Remainder. I.e.,	
			\mydef{Classical Hypotheses set of a program}{cHyps}{
				Let $P$ be an NLP.
				We write $CHyps(P)$ to denote the set of assumable hypotheses of $P$ consistent with the classical notion of support: the atoms that
				appear as DNLs in the bodies of rules of $\widehat{P}$.
				Formally, $CHyps(P)=\{a:\exists_{r\in \widehat{P}}\dnot a\in body(r)\}$.
			}
			Here we take the layered support compatible approach and, therefore, we will use the Hypotheses set as in definition
			\ref{def:hyps}.
			Since $CHyps(P)\subseteq Hyps(P)$ for every NLP $P$, there is no generality loss in using $Hyps(P)$ instead of $CHyps(P)$, while
			using $Hyps(P)$ allows for some useful semantics properties examined in the sequel.		
		\subsection{Definition}
	%
			Intuitively, a \MH\ \m\ of a program is obtained from a minimal set of hypotheses which is 
			sufficiently large to determine the truth-value of all literals via Remainder.
			\mydef{\MH\ \m}{MHm}{
				Let $P$ be an NLP. 
				Let $Hyps(P)$ be the set of assumable hypotheses of $P$ (cf. definition \ref{def:hyps}), and $H$ some subset of $Hyps(P)$.
				
				A \twov\ \m\ $M$ of $P$ is a \MH\ \m\ of $P$ iff $$M^{+}=facts(\widehat{P\cup H})=heads(\widehat{P\cup H})$$
				where 
				$H=\emptyset$ or 
				$H$ is non-empty set-inclusion minimal (the set-inclusion minimality is considered only for non-empty $H$s).
				I.e., the hypotheses set $H$ is minimal but sufficient to determine (via Remainder) the truth-value of all literals in the program.
			}
			We already know that $WFM^+(P)=facts(\widehat{P})$ and that $WFM^{+u}(P)=heads(\widehat{P})$.
			Thus, whenever $facts(\widehat{P})=heads(\widehat{P})$ we have $WFM^+(P)=WFM^{+u}(P)$ which means $WFM^u(P)=\emptyset$.
			Moreover, whenever $WFM^u(P)=\emptyset$ we know, by Corollary 5.6 of \cite{WFS}, that the \twov\ \m\ $M$ such that 
			$M^+=facts(\widehat{P})$ is the unique \sm\ of $P$.
			Thus, we conclude that, as an alternative equivalent definition, $M$ is a \MH\ \m\ of $P$ iff $M$ is a \sm\ of $P\cup H$ where $H$ is 
			empty or a non-empty set-inclusion minimal subset of $Hyps(P)$.
			Moreover, it follows immediately that every SM of $P$ is a \MH\ \m\ of $P$.			
	
				In example \ref{ex:LRem(P)vsRem(P)} we can thus see that we have the two \ms\ $\{beach, mountain,\\ \dnot travel\}$ and 
			$\{travel, beach, \dnot mountain\}$.
			This is the case because the addition of the fourth stubborn friend does not change the set of $Hyps(P)$ which is based upon the Layered
			Remainder, and not on the Remainder.
			
		\amp{
		\myex{\MH\ \ms\ for the vacation with passport variation}{MHVacationPassport}{
			Consider again the vacation problem from example \ref{ex:vacation} with a variation including the need for valid passports for travelling 
			$P=$
			\myprog{
				beach & \< & \dnot mountain\\
				mountain & \< & \dnot travel\\
				travel & \< & \dnot beach, \dnot expired\_passport\\
				&&\\
				passport\_ok & \< & \dnot expired\_passport\\
				expired\_passport & \< & \dnot passport\_ok
			}
			We have $P=\mathring{P}=\widehat{P}$ and thus $Hyps(P)=\{beach, mountain, travel, passport\_ok,expired\_passport\}$.
			Let us see which are the MH \ms\ for this program.\\ 
			$H=\emptyset$ does not yield a MH \m.\\
			Assuming $H=\{beach\}$ we have $P\cup H=P\cup\{beach\}=$
			\myprog{
				beach & \< & \dnot mountain\\
				mountain & \< & \dnot travel\\
				travel & \< & \dnot beach, \dnot expired\_passport\\
				beach&&\\
				passport\_ok & \< & \dnot expired\_passport\\
				expired\_passport & \< & \dnot passport\_ok
			}
			and $\widehat{P\cup H}=$
			\myprog{
				mountain&&\\
				beach&&\\
				passport\_ok & \< & \dnot expired\_passport\\
				expired\_passport & \< & \dnot passport\_ok
			}
			which means $H=\{beach\}$ is not sufficient to determine the truth values of all literals of $P$.
			One can easily see that the same happens for $H=\{mountain\}$ and for $H=\{travel\}$: in either case the literals $passport\_ok$ and
			$expired\_passport$ remain non-determined.\\			
			If we assume $H=\{expired\_passport\}$ then $P\cup H$ is
			\myprog{
				beach & \< & \dnot mountain\\
				mountain & \< & \dnot travel\\
				travel & \< & \dnot beach, \dnot expired\_passport\\
				&&\\
				passport\_ok & \< & \dnot expired\_passport\\
				expired\_passport & \< & \dnot passport\_ok\\
				expired\_passport &&
			}
			and $\widehat{P\cup H}=$
			\myprog{
				mountain &&\\
				expired\_passport &&
			}
			which means 
			$M_{expired\_passport}^+=facts(\widehat{P\cup H})=heads(\widehat{P\cup H})=\{mountain,expired\_passport\}$, i.e., \\
			$M_{expired\_passport}=\{\dnot beach,mountain,\dnot travel, \dnot passport\_ok,expired\_passport\}$, is a MH \m\ of $P$.
			Since assuming $H=\{expired\_passport\}$ alone is sufficient to determine all literals, there is no other set of hypotheses $H'$ of $P$
			such that $H'\supset\{expired\_passport\}$ (notice the \emph{strict} $\supset$, not $\supseteq$), yielding a MH \m\ of $P$.
			E.g., $H'=\{travel,expired\_passport\}$ does not lead to a MH \m\ of $P$ simply because $H'$ is not minimal w.r.t.
			$H=\{expired\_passport\}$.\\			
			If we assume $H=\{passport\_ok\}$ then $P\cup H$ is
			\myprog{
				beach & \< & \dnot mountain\\
				mountain & \< & \dnot travel\\
				travel & \< & \dnot beach, \dnot expired\_passport\\
				&&\\
				passport\_ok & \< & \dnot expired\_passport\\
				expired\_passport & \< & \dnot passport\_ok\\
				passport\_ok &&
			}
			and $\widehat{P\cup H}=$
			\myprog{
				beach & \< & \dnot mountain\\
				mountain & \< & \dnot travel\\
				travel & \< & \dnot beach\\
				passport\_ok &&
			}
			which, apart from the fact $passport\_ok$, corresponds to the original version of this example and still leaves literals with non-determined
			truth-values.
			I.e., assuming the passports are OK allows for the three possibilities of example \ref{ex:vacation} but it is not enough to entirely ``solve''
			the vacation problem: we need some hypotheses set containing one of $beach$, $mountain$, or $travel$ if (in this case, 
			\emph{and only if}) it also contains $passport\_ok$.
		}
		}
			
		\amp{
		\myex{Minimality of Hypotheses does not guarantee minimality of \m}{MHnotGuaranteesMM}{
			Let $P$, with no SMs, be
			\myprog{
				a &\<& \dnot b, c\\
				b &\<& \dnot c, \dnot a\\
				c &\<& \dnot a, b
			}
			In this case $P=\widehat{P}=\mathring{P}$, which makes $Hyps(P)=\{a,b,c\}$.\\			
			$H=\emptyset$ does not determine all literals of $P$ because
			$facts(\widehat{P\cup\emptyset})=facts(\widehat{P})=\emptyset$ and
			$heads(\widehat{P\cup\emptyset})=heads(\widehat{P})=\{a,b,c\}$.\\
			$H=\{a\}$ does determine all literals of $P$ because
			$facts(\widehat{P\cup\{a\}})=\{a\}$ and 
			$heads(\widehat{P\cup\{a\}})=\{a\}$,
			thus yielding the MH \m\ $M_{a}$ such that $M_{a}^{+}=facts(\widehat{P\cup\{a\}})=\{a\}$, i.e., $M_{a}=\{a,\dnot b,\dnot c\}$.\\			
			$H=\{c\}$ is also a minimal set of hypotheses determining all literals because
			$facts(\widehat{P\cup\{c\}})=\{a,c\}$ and 
			$heads(\widehat{P\cup\{c\}})=\{a,c\}$,
			thus yielding the MH \m\ $M_{c}$ of $P$ such that $M_{c}^{+}=facts(\widehat{P\cup\{c\}})=\{a,c\}$, i.e., $M_{c}=\{a,\dnot b,c\}$.
			However, $M_{c}$ is not a \mm\ of $P$ because $M_{c}^{+}=\{a,c\}$ is a strict superset of $M_{a}^{+}=\{a\}$.
			$M_{c}$ is indeed an MH \m\ of $P$, but just not a minimal \m\ thereby being a clear example of how minimality of hypotheses does not 
			entail minimality of consequences.
			Just to make this example complete, we show that $H=\{b\}$ also determines all literals of $P$ because
			$facts(\widehat{P\cup\{b\}})=\{b,c\}$ and
			$heads(\widehat{P\cup\{b\}})=\{b,c\}$, 
			thus yielding the MH \m\ $M_{b}$ such that $M_{b}^{+}=facts(\widehat{P\cup\{b\}})=\{b,c\}$, i.e., $M_{b}=\{\dnot a, b, c\}$.
			Any other hypotheses set is necessarily a strict superset of either $H=\{a\}$, $H=\{b\}$, or $H=\{c\}$ and, therefore, not set-inclusion minimal; i.e., there are no more MH \ms\ of $P$.
		}
		Also, not all \mms\ of a program are MH \ms, 
		as the following example shows.
		\myex{Some \mms\ are not Minimal Hypotheses \ms}{LDm=/=>MHm}{
			Let $P$ (with no SMs) be
			\myprog{
				a & \< & k\\
				k & \< & \dnot t\\
				t & \< & a, b\\
				a & \< & \dnot b\\
				b & \< & \dnot a
			}
			In this case $P=\widehat{P}=\mathring{P}$ and therefore $Hyps(P)=\{a,b,t\}$.
			Since $facts(\widehat{P})\neq heads(\widehat{P})$, the hypotheses set $H=\emptyset$ does not yield a MH \m.		
			Assuming $H=\{a\}$ we have $\widehat{P\cup H}=\widehat{P\cup\{a\}}=\{a\<\textrm{\ \ \ ,\ \ \ } k\<\}$
			so, $\widehat{P\cup H}$ is the set of facts $\{a,k\}$ and, therefore, $M_{a}$ such that 
			$M_{a}^{+}=facts(\widehat{P\cup H})=facts(\widehat{P\cup\{a\}})=\{a,k\}$, is a MH \m\ of $P$.
			Assuming $H=\{b\}$ we have $\widehat{P\cup\{b\}}=$
			\myprog{
				a &\<& k\\
				k &\<& \dnot t\\
				t &\<& a\\
				b &\<& \dnot a\\
				b&&
			}
			thus $facts(\widehat{P\cup\{b\}})=\{b\}\neq heads(\widehat{P\cup\{b\}})=\{a,b,t,k\}$, which means the set of hypotheses $H=\{b\}$ does
			not yield a MH \m\ of $P$.
			Assuming $H=\{t\}$ we have $\widehat{P\cup\{t\}}=$
			\myprog{
				t &\<& a,b\\
				b &\<& \dnot a\\
				a &\<& \dnot b\\
				t&&
			}
			thus $facts(\widehat{P\cup\{t\}})=\{t\}\neq heads(\widehat{P\cup\{t\}})=\{a,b,t\}$, which means the set of hypotheses $H=\{t\}$ does not
			yield a MH \m\ of $P$.\\			
			Since we already know that $H=\{a\}$ yields an MH \m\ $M_{a}$ with $M_{a}^{+}=\{a,k\}$, there is no point in trying out any subset $H'$ 
			of $Hyps(P)=\{a,b,t\}$ such that $a\in H'$ because any such subset would not be minimal w.r.t. $H=\{a\}$.
			Let us, therefore, move on to the unique subset left: $H=\{b,t\}$.
			Assuming $H=\{b,t\}$ we have $\widehat{P\cup\{b,t\}}=\{t\<\textrm{\ \ \ ,\ \ \ } b\<\}$
			thus $facts(\widehat{P\cup\{b,t\}})=\{b,t\}=heads(\widehat{P\cup\{b,t\}})$, which means $M_{b,t}$ such that 
			$M_{b,t}^{+}=facts(\widehat{P\cup H})=facts(\widehat{P\cup\{b,t\}})=\{b,t\}$, is a MH \m\ of $P$.\\			
			It is important to remark that this program has other \cms, e.g, $\{a,k\}, \{b,t\}$, and $\{a,t\}$, 
			but only the first two are \MH\ \ms\ --- $\{a,t\}$ is obtainable only via the set of hypotheses $\{a,t\}$ which is non-minimal w.r.t.
			$H=\{a\}$ that yields the MH \m\ $\{a,k\}$.
		}
		}

		\subsection{Properties}
			The minimality of $H$ is not sufficient to ensure minimality of $M^{+} = facts(\widehat{P\cup H})$ making its checking explicitly necessary
			if that is so desired.
			Minimality of hypotheses is indeed the common practice is science, not the minimality of their inevitable consequences.
			To the contrary, the more of these the better because it signifies a greater predictive power.
	
			In Logic Programming \m\ minimality is a consequence of definitions: the $T$ operator in definite programs is conducive to
			defining a
			least fixed point, a unique \mm\ semantics; in SM, though there may be more than one \m, minimality turns out to be a property because
			of the stability (and its attendant classical support) requirement; in the WFS, again the existence of a least fixed point operator affords a
			minimal (information) \m.
			In abduction too, minimality of consequences is not a caveat, but rather minimality of hypotheses is, if that even.
			Hence our approach to LP semantics via MHS is novel indeed, and insisting instead on positive hypotheses establishes an improved and
			more general link to abduction and argumentation \cite{Pereira:2007fv,lmp:arg07}.
			\mytheo{At least one Minimal Hypotheses \m\ of $P$ complies with the \WFM}{ExistsMH>=WFM}{
				Let $P$ be an NLP.
				Then, there is at least one Minimal Hypotheses \m\ $M$ of $P$ such that 
				$M^{+}\supseteq WFM^{+}(P)$ and $M^{+}\subseteq WFM^{+u}(P)$.
			}{
				If $facts(\widehat{P})=heads(\widehat{P})$ or equivalently, $WFM^u(P)=\emptyset$, then $M_H$ is a MH \m\
				of $P$ given that $H=\emptyset$ because
				$M_H^+=facts(\widehat{P\cup H})=heads(\widehat{P\cup H})=facts(\widehat{P\cup\emptyset})=heads(\widehat{P\cup\emptyset})=facts(\widehat{P})=heads(\widehat{P})$.
				On the other hand, if $facts(\widehat{P})\neq heads(\widehat{P})$, then there is at least one non-empty set-inclusion minimal set of hypotheses $H\subseteq Hyps(P)$
				such that $H\supseteq facts(P)$.
				The corresponding $M_H$ is, by definition, a MH \m\ of $P$ which is guaranteed to comply with $M_H^+\supseteq WFM^+(P)=facts(\widehat{P})$
				and $M_H^-\supseteq\dnot WFM^-(P)=\dnot(\mathcal{H}_P\setminus M_H^+)$.
			}
			\mytheo{Minimal Hypotheses semantics guarantees \m\ existence}{MHModelExistence}{
				Let $P$ be an NLP.
				There is always, at least, one Minimal Hypotheses \m\ of $P$.
			}{
				It is trivial to see that one can always find a set $H\subseteq Hyps(P)$ such that
				$M_{H'}^+=facts(\widehat{P\cup H'})=heads(\widehat{P\cup H'})$ --- in the extreme case, $H'=Hyps(P)$.
				From such $H'$ one can always select a minimal subset $H\subset H$ such that
				$M_{H}^+=facts(\widehat{P\cup H})=heads(\widehat{P\cup H})$ still holds.
			}
			\subsection{Relevance}
				\mytheo{Minimal Hypotheses semantics enjoys Relevance}{MHRelevance}{
					Let $P$ be an NLP.
					Then, by definition \ref{def:relevance}, it holds that
					$$(\forall_{M\in Models_{MH}(P)}a \in M^{+}) \Leftrightarrow (\forall_{M_{a}\in Models_{MH}(Rel_{P}(a))}a \in M_{a}^{+})$$
				}{
					$\Rightarrow$:
					Assume $\forall_{M\in Models_{MH}(P)}a \in M^{+}$.
					Now we need to prove \\$\forall_{M_{a}\in Models_{MH}(Rel_{P}(a))}a \in M_{a}^{+}$.
					Assume some $M_a\in Models_{MH}(Rel_P(a))$; now we show that assuming $a\notin M_a^+$ leads to an absurdity.
					Since $M_{a}$ is a \twov\ complete \m\ of $Rel_{P}(a)$ we know that $|M_{a}|=\mathcal{H}_{Rel_{P}(a)}$ hence, if 
					$a\notin M_{a}$, then necessarily $\dnot a\in M_{a}^{-}$.
					Since $P\supseteq Rel_{P}(a)$, by theorem \ref{theo:MHModelExistence} we know that there is some \m\ $M'$ of $P$
					such that $M'\supseteq M_a$, and thus $\dnot a\in M'^-$ which contradicts the initial assumption that 
					$\forall_{M\in Models_{MH}(P)}a \in M^{+}$.
%
%
%
					We conclude $a\notin M_{a}$ cannot hold, i.e., $a\in M_{a}$ must hold.
					Since $a \in M^{+}$ hold for every \m\ $M$ of $P$, then $a\in M_{a}$ must hold for every \m\ $M_a$ of $Rel_{P}(a)$.
					
					$\Leftarrow$:
					Assume $\forall_{M_{a}\in Models_{MH}(Rel_{P}(a))}a \in M_{a}^{+}$.
					Now we need to prove \\$\forall_{M\in Models_{MH}(P)}a \in M^{+}$.
					Let us write $P_{)a(}$ as an abbreviation of $P\setminus Rel_{P}(a)$.
					We have therefore $P=P_{)a(}\cup Rel_{P}(a)$.
					Let us now take $P_{)a(}\cup M_{a}$.
					We know that every NLP as an MH \m, hence every MH \m\ $M$ of $P_{)a(}\cup M_{a}$ is such that $M\supseteq M_{a}$.
					Let $H_{M_{a}}$ denote the Hypotheses set of $M_{a}$ --- i.e.,
					$M_{a}^+=facts(\widehat{Rel_{P}(a)\cup H_{M_{a}}})=heads(\widehat{Rel_{P}(a)\cup H_{M_{a}}})$, with $H_{M_{a}}=\emptyset$ or 
					non-empty set-inclusion minimal, as per definition \ref{def:MHm}.
					If $facts(\widehat{P\cup H_{M_{a}}})=heads(\widehat{P\cup H_{M_{a}}})$ then 
					$M^+=facts(\widehat{P\cup H_{M}})=heads(\widehat{P\cup H_{M}})$ is an MH \m\ of $P$ with $H_{M}=H_{M_{a}}$ and, necessarily, 
					$M\supseteq M_{a}$.
		
					If $facts(\widehat{P\cup H_{M_{a}}})\neq heads(\widehat{P\cup H_{M_{a}}})$ then, knowing that every program has a MH \m, we can
					always find an MH \m\ $M$ of $P_{)a(}\cup M_{a}$, with $H'\subseteq Hyps(P_{)a(}\cup M_{a})$, where 
					$M^+=facts(\widehat{P\cup H'})=heads(\widehat{P\cup H'})$.
					Such $M$ is thus $M^+=facts(\widehat{P\cup H_M})=heads(\widehat{P\cup H_M})$ where
					$H_M=H_{M_a}\cup H'$, which means $M$ is a MH \m\ of $P$ with $M\supseteq M_a$.
					Since every \m\ $M_a$ of $Rel_P(a)$ is such that $a\in M_a^+$, then every \m\ $M$ of $P$ must also be such that $a\in M$.
				}
			\subsection{Cumulativity}
				MH semantics enjoys Cumulativity 
				thus allowing for lemma storing techniques to be used during
				computation of answers to queries.
				\mytheo{Minimal Hypotheses semantics enjoys Cumulativity}{MHCumulativity}{
					Let $P$ be an NLP.
					Then
					\vspace{-3mm}
					\begin{center}
						$ \forall_{a,b\in\mathcal{H}_{\mathcal{P}}}\big((\forall_{M\in Models_{MH}(\mathcal{P})} a\in M^{+})\Rightarrow $\\
						$(\forall_{M\in Models_{MH}(\mathcal{P})} b\in M^{+}\Leftrightarrow 
							\forall_{M_{a}\in Models_{MH}(\mathcal{P}\cup\{a\})} b\in M_{a}^{+})\big)$
					\end{center}
				}{
					Assume $\forall_{\substack{a\in\mathcal{H}_{\mathcal{P}}\\M\in Models_{MH}(\mathcal{P})}} a\in M^{+}$.
					
					$\Rightarrow$:
					Assume $\forall_{M\in Models_{MH}(\mathcal{P})} b\in M^{+}$.
					Since every MH \m\ $M$ contains $a$ it follows that all such $M$ are also MH \ms\ of $P\cup\{a\}$.
					Since we assumed $b\in M$ as well, and we know that $M$ is a MH \m\ of $P\cup\{a\}$ we conclude $b$ is also in those MH \ms\
					$M$ of $P\cup\{a\}$.
					By adding $a$ as a fact we have necessarily $Hyps(P\cup\{a\})\subseteq Hyps(P)$ which means that there cannot be more MH \ms\
					for $P\cup\{a\}$ than for $P$.
					Since we already know that for every MH \m\ $M$ of $P$, $M$ is also a MH \m\ of $P\cup\{a\}$ we must conclude that
					$\forall_{M\in Models_{MH}(P)}\exists^1_{M'\in Models_{MH}(P\cup\{a\})}$ such that $M'^+\supseteq M^+$.
					Since $\forall_{M\in Models_{MH}(\mathcal{P})} b\in M^{+}$ we necessarily conclude
					$\forall_{M_{a}\in Models_{MH}(\mathcal{P}\cup\{a\})} b\in M_{a}^{+}$.
					
					$\Leftarrow$:
					Assume $\forall_{M_{a}\in Models_{MH}(\mathcal{P}\cup\{a\})} b\in M_{a}^{+}$.
					Since the MH semantics is relevant (theorem \ref{theo:MHRelevance}) if $b$ does not depend on $a$ then adding $a$ as a fact to 
					$P$ or not has no impact on $b$'s truth-value, and if $b\in M_{a}^+$ then $b\in M^+$ as well.
					If $b$ does depend on $a$, which is true in every MH \m\ $M$ of $P$, then either 
					1) $b$ depends positively on $a$, and in this case since $a\in M$ then $b\in M$ as well; or
					2) $b$ depends negatively on $a$, and in this case the lack of $a$ as a fact in $P$ can only contribute, if at all, to make $b$ true in
						$M$ as well.	
					Then we conclude $\forall_{M\in Models_{MH}(\mathcal{P})} b\in M^{+}$.		
				}	
			\subsection{Complexity}
				The complexity issues usually relate to a particular set of tasks, namely: 
				1) knowing if the program has a \m; 
				2) if it has any \m\ entailing some set of ground literals (a query); 
				3) if all \ms\ entail a set of literals.
				In the case of MH semantics, the answer to the first question is an immediate ``yes'' because MH semantics guarantees \m\ existence
				for NLPs; the second and third questions correspond (respectively) to Brave and Cautious Reasoning, which we now analyse.
				\subsubsection{Brave Reasoning}
					The complexity of the Brave Reasoning task with MH semantics, i.e., finding an MH \m\ satisfying some particular set of literals is $\Sigma_2^P$-complete.

					\mytheo{Brave Reasoning with MH semantics is $\Sigma_2^P$-complete}{MHBraveReasoningIsSigma2P}{
						Let $P$ be an NLP, and $Q$ a set of literals, or \emph{query}.
						Finding an MH \m\ such that $M\supseteq Q$ is a $\Sigma_2^P$-complete task.
					}{
						To show that finding a MH \m\ $M\supseteq Q$ is $\Sigma_2^P$-complete, note that a nondeterministic Turing machine with access to an NP-complete oracle can solve the problem as follows: nondeterministically guess a set $H$ of hypotheses (i.e., a subset of $Hyps(P)$). It remains to check if $H$ is empty or non-empty minimal such that $M^+=facts(\widehat{P\cup H})=heads(\widehat{P\cup H})$ and $M\supseteq Q$. Checking that $M^+=facts(\widehat{P\cup H})=heads(\widehat{P\cup H})$ can be done in polynomial time (because computing $\widehat{P\cup H}$ can be done in polynomial time \cite{DBLP:journals/tplp/BrassDFZ01} for whichever $P\cup H$), and checking $H$ is empty or non-empty minimal requires a nondeterministic guess of a strict subset $H'$ of $H$ and then a polynomial check if $facts(\widehat{P\cup H'})=heads(\widehat{P\cup H'})$.
					}
										
				\subsubsection{Cautious Reasoning}
					Conversely, the Cautious Reasoning, i.e., guaranteeing that every MH \m\ satisfies some particular set of literals, is $\Pi_2^P$-complete.
					\mytheo{Cautious Reasoning with MH semantics is $\Pi_2^P$-complete}{MHCautiousReasoningIsPi2P}{
						Let $P$ be an NLP, and $Q$ a set of literals, or \emph{query}.
						Guaranteeing that all MH \ms\ are such that $M\supseteq Q$ is a $\Pi_2^P$-complete task.
					}{
						Cautious Reasoning is the complement of Brave Reasoning, and since the latter is $\Sigma_2^P$-complete
						(theorem \ref{theo:MHBraveReasoningIsSigma2P}), the former must necessarily be $\Pi_2^P$-complete.
					}					

		The set of hypotheses $Hyps(P)$ is obtained from $\mathring{P}$ which identifies rules that depend on themselves.
		The hypotheses are the atoms of DNLs of $\mathring{P}$, i.e., the ``atoms of \emph{not}s in loop''.
		A \MH\ \m\ is then obtained from a minimal set of these hypotheses sufficient to determine the \twov\ truth-value of every literal in the program.
		The MH semantics imposes no ordering or preference between hypotheses --- only their set-inclusion minimality.
		For this reason, we can think of the choosing of a set of hypotheses yielding a MH \m\ as finding a minimal solution to a disjunction problem, where the disjuncts are the hypotheses.
		In this sense, it is therefore understandable that the complexity of the reasoning tasks with MH semantics is in line with that of, e.g., reasoning tasks with SM semantics with Disjunctive \LPs, i.e, $\Sigma_2^P$-complete and $\Pi_2^P$-complete.

					\amp{
					In abductive reasoning (as well as in Belief Revision) one does not always require minimal solutions.
					Likewise, taking a hypotheses assumption based semantic approach, like the one of MH, one may not require minimality of assumed
					hypotheses.
					In such case, we would be under a non-Minimal Hypotheses semantics, and the complexity classes of the corresponding reasoning
					task would be one level down in the Polynomial Hierarchy relatively to the MH semantics, i.e., Brave Reasoning with a non-Minimal
					Hypotheses semantics would be NP-complete, and Cautious Reasoning would be coNP-complete.
					We leave the exploration of such possibilities for future work.
					}

		\subsection{Comparisons}
			As we have seen 
			all \sms\ are MH \ms.
			Since MH \ms\ are always guaranteed to exist for every NLP (cf. theorem \ref{theo:MHModelExistence}) and SMs are not, it follows
			immediately that the Minimal Hypotheses semantics is a strict \m\ conservative generalization 
			of the \SMss.
			The MH \ms\ that are \sms\ are exactly those in which all rules are classically supported.
			With this criterion one can conclude whether some program does not have any \sms.
			For \NLPs, the \SMss\ coincides with the Answer-Set semantics (which is a generalization of SMs to \ELPs), where the latter is known 
			(cf. \cite{LPsWithCNeg}) to correspond to Reiter's default logic.
			Hence, all Reiter's default extensions have a corresponding Minimal Hypotheses \m.
			Also, since Moore's expansions of an autoepistemic theory \cite{2781} are known to have a one-to-one correspondence with the \sms\ of
			the NLP version of the theory, we conclude that for every such expansion there is a matching Minimal Hypotheses \m\ for the same NLP.
			
			\amp{
			Disjunctive Logic Programs (DisjLPs --- allowing for disjunctions in the heads of rules) can be syntactically transformed into NLPs by
			applying the Shifting Rule presented in \cite{Dix96reducingdisjunctive} in all possible ways.
			By non-deterministically applying such transformation in all possible ways, several SCCs of rules may appear in the resulting NLP that
			were not present in the original DisjLP --- assigning a meaning to every such SCC is a distinctive feature of MH semantics, unlike other
			semantics such as the SMs.
			This way, the MH semantics can be defined for DisjLPs as well: the MH \ms\ of a DisjLP are the MH \ms\ of the NLP resulting from the
			transformation via Shifting Rule.
			}
			
			\amp{
			There are other kinds of disjunction, like the one in logic programs with ordered disjunction (LPOD) \cite{Brewka02logicprogramming}.
			These employ ``\emph{a new connective called ordered disjunction.
			The new connective allows to represent alternative, ranked options for problem solutions in the heads of rules}''.
			As the author of \cite{Brewka02logicprogramming} says ``the semantics of logic programs with ordered disjunction is based on a
			preference relation on answer sets.''
			This is different from the semantics assigned by MH since the latter includes no ordering, nor preferences, in the assumed minimal
			sets of hypotheses.
			E.g., in example \ref{ex:vacation} there is no notion of preference or ordering amongst candidate \ms\ --- LPODs would not be the 
			appropriate formalism for such cases.
			We leave for future work a thorough comparison of these approaches, namely comparing the semantics of LPODs against the MH \ms\ of
			LPODs transformed into NLPs (via the Shifting Rule).
			}
			
			\amp{
			The motivation for \cite{Witteveen:1995} is similar to our own --- to assign a semantics to \emph{every} NLP --- however their approach
			is different from ours in the sense that the methods in \cite{Witteveen:1995} resort to contrapositive rules allowing any positive literal in
			the head to be shifted (by negating it) to the body or any negative literal in the body to be shifted to the head (by making it positive).
			This approach considers each rule as a disjunction making no distinction between such literals occurring in the rule, whether or not they
			are in loop with the head of the rule.
			This permits the shifting operations in \cite{Witteveen:1995} to create support for atoms that have no rules in the original program.
			E.g.
			\myex{Nearly-\SMs\ vs MH \ms}{nearlySMsVsMHms}{
				Take the program
				$P=$
				\myprog{
					a&\<&\dnot b\\
					b&\<&\dnot c\\
					c&\<&\dnot a,\dnot x
				}
				According to the shifting operations in \cite{Witteveen:1995} this program could be transformed into
				$P'=$
				\myprog{
					b&\<&\dnot a\\
					b&\<&\dnot c\\
					x&\<&\dnot a,\dnot c
				}
				by
				shifting $a$ and $\dnot b$ in the first rule, and 
				shifting the $\dnot x$ to the head (becoming positive $x$) and $c$ to the body (becoming negative $\dnot c$) of the third rule
				thus allowing for $\{b,x\}$ (which is a \sm\ of $P'$) to be a \emph{nearly \sm} of $P$.
				In this sense the approach of \cite{Witteveen:1995} allows for the violation of the Closed-World Assumption.
				This does not happen with our approach: 
				$\{b,x\}$ is not a \MHm\ simply because since $x$ has no rules in $P$ it cannot be true in any MH \m\ ---
				$\dnot x$ is not a member of $Hyps(P)$ (cf. def. \ref{def:hyps}).
			}			
			}			
			
			As shown in theorem \ref{theo:ExistsMH>=WFM}, at least one MH \m\ of a program complies with its \wfm, although not necessarily all
			MH \ms\ do.
			E.g., the program in Ex. \ref{ex:LRem(P)vsRem(P)} has the two MH \ms\ $\{beach,\\mountain,\dnot travel\}$ and 
			$\{beach,\dnot mountain,travel\}$, whereas the $WFM(P)$imposes $WFM^{+}(P)\\=\{beach,mountain\}$, $WFM^{u}(P)=\emptyset$, and
			$WFM^{-}(P)=\{travel\}$.
			This is due to the set of Hypotheses $Hyps(P)$ of $P$ being taken from $\mathring{P}$ (based on the layered
			support notion) instead of being taken from $\widehat{P}$ (based on the classical notion of support).
			
			Not all Minimal Hypotheses \ms\ are \MMs\ of a program.
			The rationale behind MH semantics is minimality of hypotheses, but not necessarily minimality of consequences, the latter being
			enforceable, if so desired, as an additional requirement, although at the expense of increased complexity.
						
			The relation between \lps\ and argumentation systems has been considered for a long time now
			(\cite{dung95acceptability}	
			amongst many others)
			and we have also taken steps to understand and further that relationship
			\cite{Pereira:2007fv,lmp:arg07,lmp_amp_oppositional_concepts_chapter}.
			Dung's Preferred Extensions \cite{dung95acceptability} are maximal sets of negative hypotheses yielding consistent \ms.
			Preferred Extensions, however, these are not guaranteed to always yield \twov\ complete \ms.
			Our previous approaches \cite{Pereira:2007fv,lmp:arg07} to argumentation have already addressed the issue of \twov\ \m\ existence
			guarantee, and the MH semantics also solves that problem by virtue of positive, instead of negative, hypotheses assumption.
					
		
	\section{Conclusions and Future Work}
		Taking a positive hypotheses assumption approach we defined the \twov\ \MHs\ for NLPs that guarantees \m\ existence, enjoys relevance
		and cumulativity, and is also a \m\ conservative generalization of the SM semantics.
		Also, by adopting positive hypotheses, we not only generalized the argumentation based approach of \cite{dung95acceptability}, but the
		resulting MH semantics lends itself naturally to abductive reasoning, it being understood as hypothesizing plausible reasons sufficient for
		justifying given observations or supporting desired goals.
		We also defined the layered support notion which generalizes the classical one by recognizing the special role of loops.
		
		For query answering, the MH semantics provides mainly three advantages over the SMs:
		1) by enjoying Relevance top-down query-solving is possible, thereby circumventing whole \m\ computation (and grounding) which is
		unavoidable with SMs;
		2) by considering only the relevant sub-part of the program when answering a query it is possible to enact grounding of only those rules, if
		grounding is really desired, whereas with SM semantics whole program grounding is, once again, inevitable --- grounding is known to be a
		major source of computational time consumption; MH semantics, by enjoying Relevance, permits curbing this task to the minimum sufficient
		to answer a query;
		3) by enjoying Cumulativity, as soon as the truth-value of a literal is determined in a branch for the top query it can be stored in a table and
		its value used to speed up the computations of other branches within the same top query.
		
		Goal-driven abductive reasoning is elegantly modelled by top-down abductive-query-solving. 
		By taking a hypotheses assumption approach, enjoying Relevance, MH semantics caters well for this convenient problem representation and
		reasoning category.
	
		Many applications have been developed using the \SM/Answer-set semantics as the underlying platform.
		These generally tend to be focused on solving problems that require complete knowledge, such as search problems where all the knowledge
		represented is relevant to the solutions.
		However, as \KBs\ increase in size and complexity, and as merging and updating of KBs becomes more and more common, e.g. for
		Semantic Web applications, \cite{MKNFpadl10},
		partial knowledge problem solving importance grows, as the need to ensure overall
		consistency of the merged/updated KBs.
		
		The \MHs\ is intended to, and can be used in
		\emph{all} the applications where the \SMs/\ASs\ semantics are themselves used to model KRR and search problems,
		\emph{plus} all applications where query answering (both under a credulous mode of reasoning and under a skeptical one) is intented, 
		\emph{plus} all applications where abductive reasoning is needed.
		The MH semantics aims to be a sound theoretical platform for \twov\ (possibly abductive) reasoning with \lps.
	
		Much work still remains to be done that can be rooted in this platform contribution.
		The general topics of using non-normal \lps\ (allowing for negation, default and/or explicit, in the heads of rules) for Belief Revision, 
		Updates, 
		Preferences, 
		etc., are \emph{per se} orthogonal to the semantics issue, and therefore, all these subjects can now be addressed with \MHs\ as the
		underlying platform.
		Importantly, MH can guarantee the liveness of updated and self-updating LP programs such as those of EVOLP \cite{EVOLP} and related
		applications.
		The \MHs\ still has to be thoroughly compared with \RSMs\ \cite{PP05}, P\SMs\ \cite{DBLP:conf/micai/OsorioN07}, and other related
		semantics.
								
		In summary, we have provided a fresh platform on which to re-examine ever present issues in Logic Programming and its uses, which
		purports to provide a natural continuation and improvement of LP development.

	\bibliographystyle{plain}
	
	\bibliography{bibliography}

\begin{thebibliography}{10}

\bibitem{EVOLP}
J.J. Alferes, A.~Brogi, J.~A. Leite, and L.~M. Pereira.
\newblock Evolving logic programs.
\newblock In S.~Flesca et~al., editor, {\em Procs. JELIA'02}, volume 2424 of
  {\em LNCS}, pages 50--61. Springer, 2002.

\bibitem{DBLP:journals/tplp/BrassDFZ01}
S.~Brass, J.~Dix, B.~Freitag, and U.~Zukowski.
\newblock Transformation-based bottom-up computation of the well-founded model.
\newblock {\em TPLP}, 1(5):497--538, 2001.

\bibitem{Brewka02logicprogramming}
Gerhard Brewka.
\newblock Logic programming with ordered disjunction.
\newblock In {\em In Proceedings of AAAI-02}, pages 100--105. AAAI Press, 2002.

\bibitem{Dix95aclassification}
J.~Dix.
\newblock {A} {C}lassification {T}heory of {S}emantics of {N}ormal {L}ogic
  {P}rograms: {I}. {S}trong {P}roperties.
\newblock {\em Fundam. Inform.}, 22(3):227--255, 1995.

\bibitem{dix93aclassificationtheoryii}
J.~Dix.
\newblock {A} {C}lassification {T}heory of {S}emantics of {N}ormal {L}ogic
  {P}rograms: {I}{I}. {W}eak {P}roperties.
\newblock {\em Fundam. Inform.}, 22(3):257--288, 1995.

\bibitem{Dix96reducingdisjunctive}
J{\"u}rgen Dix, J~Urgen Dix, Georg Gottlob, Wiktor Marek, and Cecylia Rauszer.
\newblock Reducing disjunctive to non-disjunctive semantics by
  shift-operations.
\newblock {\em Fundamenta Informaticae}, 28:87--100, 1996.

\bibitem{dung95acceptability}
P.~M. Dung.
\newblock On the acceptability of arguments and its fundamental role in
  nonmonotonic reasoning, logic programming and n-person games.
\newblock {\em AI}, 77(2):321--358, 1995.

\bibitem{WFS}
A.~Van Gelder, K.~A. Ross, and J.~S. Schlipf.
\newblock The well-founded semantics for general logic programs.
\newblock {\em J. ACM}, 38(3):620--650, 1991.

\bibitem{SM-GL}
M.~Gelfond and V.~Lifschitz.
\newblock The stable model semantics for logic programming.
\newblock In {\em Procs. ICLP'88}, pages 1070--1080, 1988.

\bibitem{LPsWithCNeg}
M.~Gelfond and V.~Lifschitz.
\newblock Logic programs with classical negation.
\newblock In D.~Warren et~al., editor, {\em ICLP}, pages 579--597. MIT Press,
  1990.

\bibitem{MKNFpadl10}
A.~S. Gomes, J.~J. Alferes, and T.~Swift.
\newblock Implementing query answering for hybrid mknf knowledge bases.
\newblock In M.~Carro et~al., editor, {\em PADL'10}, volume 5937 of {\em LNCS},
  pages 25--39. Springer, 2010.

\bibitem{DBLP:journals/logcom/KakasKT92}
A.~C. Kakas, R.~A. Kowalski, and F.~Toni.
\newblock Abductive logic programming.
\newblock {\em J. Log. Comput.}, 2(6):719--770, 1992.

\bibitem{2781}
R.~C. Moore.
\newblock Semantical considerations on nonmonotonic logic.
\newblock {\em AI}, 25(1):75--94, 1985.

\bibitem{DBLP:conf/micai/OsorioN07}
M.~Osorio and J.~C. Nieves.
\newblock Pstable semantics for possibilistic logic programs.
\newblock In {\em MICAI'07}, volume 4827 of {\em LNCS}, pages 294--304.
  Springer, 2007.

\bibitem{PP05}
L.~M. Pereira and A.~M. Pinto.
\newblock Revised stable models - a semantics for logic programs.
\newblock In C.~Bento et~al., editor, {\em Procs. EPIA'05}, volume 3808 of {\em
  LNAI}, pages 29--42. Springer, 2005.

\bibitem{Pereira:2007fv}
L.~M. Pereira and A.~M. Pinto.
\newblock Approved models for normal logic programs.
\newblock In N.~Dershowitz and A.~Voronkov, editors, {\em Procs. LPAR'07},
  volume 4790 of {\em LNAI}. Springer, 2007.

\bibitem{lmp:arg07}
L.~M. Pereira and A.~M. Pinto.
\newblock Reductio ad absurdum argumentation in normal logic programs.
\newblock In G.~Simari et~al., editor, {\em ArgNMR'07-LPNMR'07}, pages 96--113.
  Springer, 2007.

\bibitem{lmp_amp_oppositional_concepts_chapter}
L.~M. Pereira and A.~M. Pinto.
\newblock {\em Collaborative vs. Conflicting Learning, Evolution and
  Argumentation, in: Oppositional Concepts in Computational Intelligence}.
\newblock Studies in Computational Intelligence 155. Springer, 2008.

\bibitem{PHDAMP}
A.~M. Pinto.
\newblock {\em Every normal logic program has a 2-valued semantics: theory,
  extensions, applications, implementations}.
\newblock PhD thesis, Universidade Nova de Lisboa, 2011.

\bibitem{citeulike:2204443}
R.~Tarjan.
\newblock Depth-first search and linear graph algorithms.
\newblock {\em SIAM J. on Computing}, 1(2):146--160, 1972.

\bibitem{Witteveen:1995}
C.~Witteveen.
\newblock Every normal program has a nearly stable model.
\newblock In J.~Dix, L.M. Pereira, and T.C. Przymusinski, editors, {\em
  Non-Monotonic Extensions of Logic Programming}, volume 927 of {\em Lecture
  Notes in Artificial Intelligence}, pages 68--84. Springer Verlag, Berlin,
  1995.

\end{thebibliography}
\end{document}